\documentclass[twocolumn,prl,tightenlines,superscriptaddress,showpacs]{revtex4-2}

\usepackage{amsmath}
\usepackage{amssymb,amsfonts,latexsym}
\usepackage{bm}
\usepackage[mathcal]{euscript}
\usepackage{graphicx}
\usepackage{epsfig}
\usepackage{color}
\usepackage{xfrac}
\usepackage{mathrsfs}
\usepackage{float}
\usepackage{hyperref}
\usepackage{url}
\usepackage{CJK}

\begin{document}

\begin{CJK*}{UTF8}{gbsn}

\title{Phase Diagram of the Two-Dimensional Vicsek Model}

\author{Yu Duan ({\CJKfamily{gbsn}段煜})}
\affiliation{Center for Soft Condensed Matter Physics and Interdisciplinary Research \& School of Physical Science and Technology, Soochow University, Suzhou 215006, China}
\affiliation{Max Planck Institute for Dynamics and Self-Organization (MPI-DS), 37077 G\"ottingen, Germany}
\affiliation{Jiangsu Physical Science Research Center, Nanjing 210093, China}

\author{Hugues Chat\'{e}}
\affiliation{Service de Physique de l'Etat Condens\'e, CEA, CNRS Universit\'e Paris-Saclay, CEA-Saclay, 91191 Gif-sur-Yvette, France}
\affiliation{Sorbonne Universit\'e, CNRS, Laboratoire de Physique Th\'eorique de la Mati\`ere Condens\'ee, 75005 Paris, France}

\author{Beno\^it Mahault}
\affiliation{Laboratoire Charles Coulomb, Universit\'e de Montpellier, CNRS, Montpellier, France}
\affiliation{Max Planck Institute for Dynamics and Self-Organization (MPI-DS), 37077 G\"ottingen, Germany}

\author{Xia-qing Shi ({\CJKfamily{gbsn}施夏清})}
\affiliation{Center for Soft Condensed Matter Physics and Interdisciplinary Research \& School of Physical Science and Technology, Soochow University, Suzhou 215006, China}

\date{\today}

\begin{abstract}
Motivated by gaps in our current knowledge of the Vicsek model, 
we present a major update of its phase diagram in two space dimensions, 
summarizing comprehensive numerical
simulations at length- and timescales never explored before. Our main conclusions are that 
multiple phase separation scenarios are at play, not just the simple microphase separation put forward in 
[Phys. Rev. Lett. 114, 068101 (2015)], and that the coexistence phase is subdivided in two.
We also map out the region of metastability of the homogeneous Toner-Tu liquid phase.
\end{abstract}

\maketitle
\end{CJK*}

The Vicsek model was devised in 1995 by Vicsek and collaborators as a minimal model for collective 
motion~\cite{vicsek1995novel}.
It is indeed simple: constant-speed self-propelled particles interact only via local alignment of their velocities 
in a noisy environment.
Since then, the Vicsek model and similar ones have taken a central role in the now-large field of active matter studies~\cite{marchetti2013hydrodynamics,bechinger2016active,tevrugtColloquiumWhatWe2026} 
and continue to be a favorite testbed for various ideas and theoretical debates, beyond the initial 
motivation (for recent examples, see~\cite{ventejou2021susceptibility,fruchart2021non-reciprocal,holubec2021finite,cavagna2021vicsek,duan2021breakdown,Chatterjee2023flocking,fava2024strong,kreienkampNonreciprocalAlignmentInduces2024,tangReentrantPhaseBehavior2025,zhao2025emergent,martin2025transition,lardet2027flocking,myinNonreciprocityGenericMechanism2026b,myinBreakdownEmergentChiral2026,wooExtensiveSpatiotemporalChaos2026}). 

The Vicsek model caught the attention of the statistical physics community early: as soon as 1995, Toner and Tu (TT) argued
that the homogeneous collective motion phase of this flying XY model has true long-range orientational order 
even in two-dimensions~ \cite{toner1995long}, 
in spite of spontaneously breaking the rotational symmetry,
a spectacular departure from equilibrium physics~\cite{tasaki2020hohenberg}.

\begin{figure}[t!]
\includegraphics[width=\columnwidth]{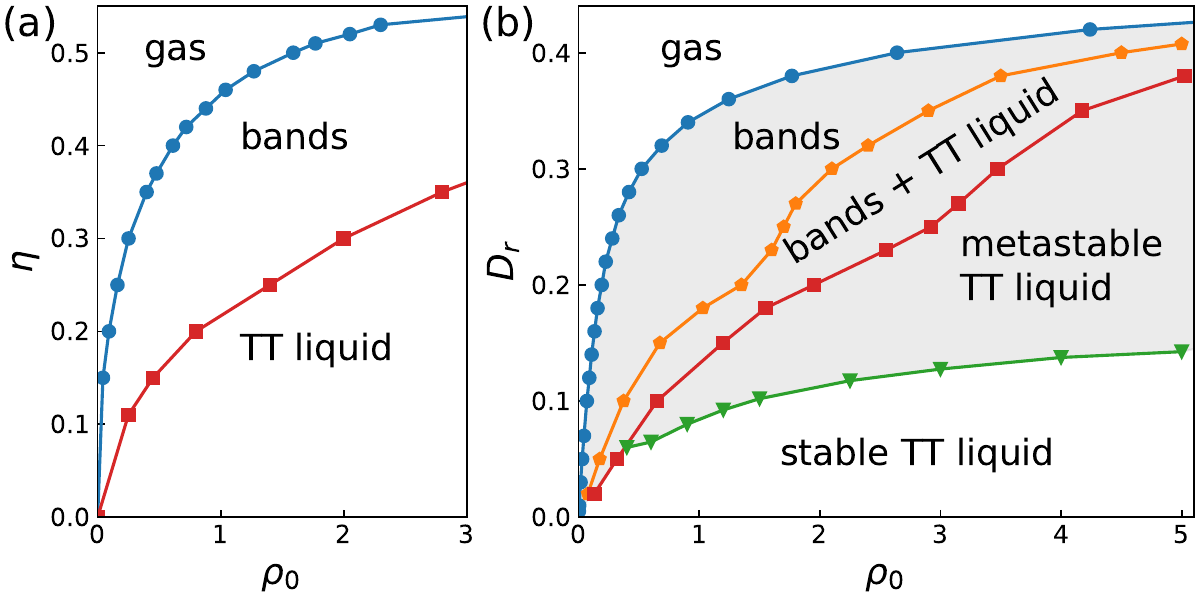}
    \caption{Phase diagram of the two-dimensional Vicsek model in the (density, noise strength) plane.
    (a) current phase diagram (discrete-time version, $v_0=0.5$, data from \cite{solon2015phase}).
    (b) new version elaborated from the results obtained here with Eq.~\eqref{VMcont}.}
    \label{fig1}
\end{figure}

It nevertheless took many years to arrive at the current basic knowledge we have of the two-dimensional 
Vicsek model, 
and in particular of its ``standard" phase diagram, usually presented in terms of the two main parameters 
at play, the global density of particles and the strength of the noise.  
It is now well known that, contrary to early claims, the order-disorder transition is not continuous, 
but occurs in two steps, in a scenario much akin to phase separation~\cite{solon2013revisiting,solon2015phase}. 
A coexistence region delimited by two binodal lines separates the disordered ``gas" phase present 
at strong noise/low density from the ordered collectively moving ``liquid'' on which Toner and Tu focused 
(Fig.~\ref{fig1}(a)). 
In this region, particles form dense, ordered bands (``microphases") that travel in the remaining sparse gas~\cite{solon2015phase}.

Some cracks have appeared in this relatively simple picture. 
(i) Whereas the coexistence phase 
was best known for exhibiting regular, smectic trains of traveling bands, K\"ursten and Ihle showed that 
they can arrange themselves, near the liquid binodal, into a regular pattern appearing as two identical smectic trains superimposed at some angle, 
and argued that these ``cross-sea"  configurations constitute a fourth phase~\cite{kursten2020dry}.
(ii) It was found that the TT liquid phase, at large-enough noise, 
can be metastable to the nucleation of growing 
counter-propagating droplets~\cite{codina2022small}, a mechanism understood analytically only recently~\cite{granek2026how}; however, the corresponding region of the phase diagram was not mapped.
(iii) To be true, the liquid binodal line separating the coexistence region from the ordered liquid was not located
 properly, relying on some operational procedure ill-defined in 
large systems (see \cite{gregoire2004onset,chate2008collective,solon2015phase,chate2020dry}). 

On top of all this, the Vicsek model was never much explored at large densities. 
The resulting situation is that it needs to be revisited. 
With this goal in mind, we embarked on a comprehensive study, drawing on modern numerical resources.

In this Letter, we present the gist of this enterprise, summarized in the new phase diagram of the 
two-dimensional Vicsek model shown in Fig.~\ref{fig1}(b), 
while our full findings are presented in 
a companion paper (CP)~\cite{CP}.
One general conclusion is that phase separation is indeed a valid concept to describe the coexistence region 
but one that appears at multiple levels. 
We find that the coexistence phase is divided in two parts, but not as K\"ursten and Ihle suggested 
in ~\cite{kursten2020dry}.
Of the simple scenario put forward in ~\cite{solon2015phase}, only the gas binodal (blue line) remains. 
We delimit a region where large domains of  ordered liquid coexist in space with band configurations, 
amounting to a ``super phase separation" scenario with its two binodals (orange and red lines). 
Moreover, at strong enough noise this super phase separation is not steady in time, with the large liquid domains
disappearing and reappearing intermittently. 
We show that the ordered liquid is partially metastable only at large-enough densities 
and locate the metastability limit (green line).
We also encountered, along the way, a large variety of configurations at moderate system sizes.
We discuss them briefly here while they are presented in detail in the CP. 
We finally argue that simple versions of continuous Toner-Tu theories cannot account for many of our findings, 
calling for new approaches.

\begin{figure}[t!]
\includegraphics[width=\columnwidth]{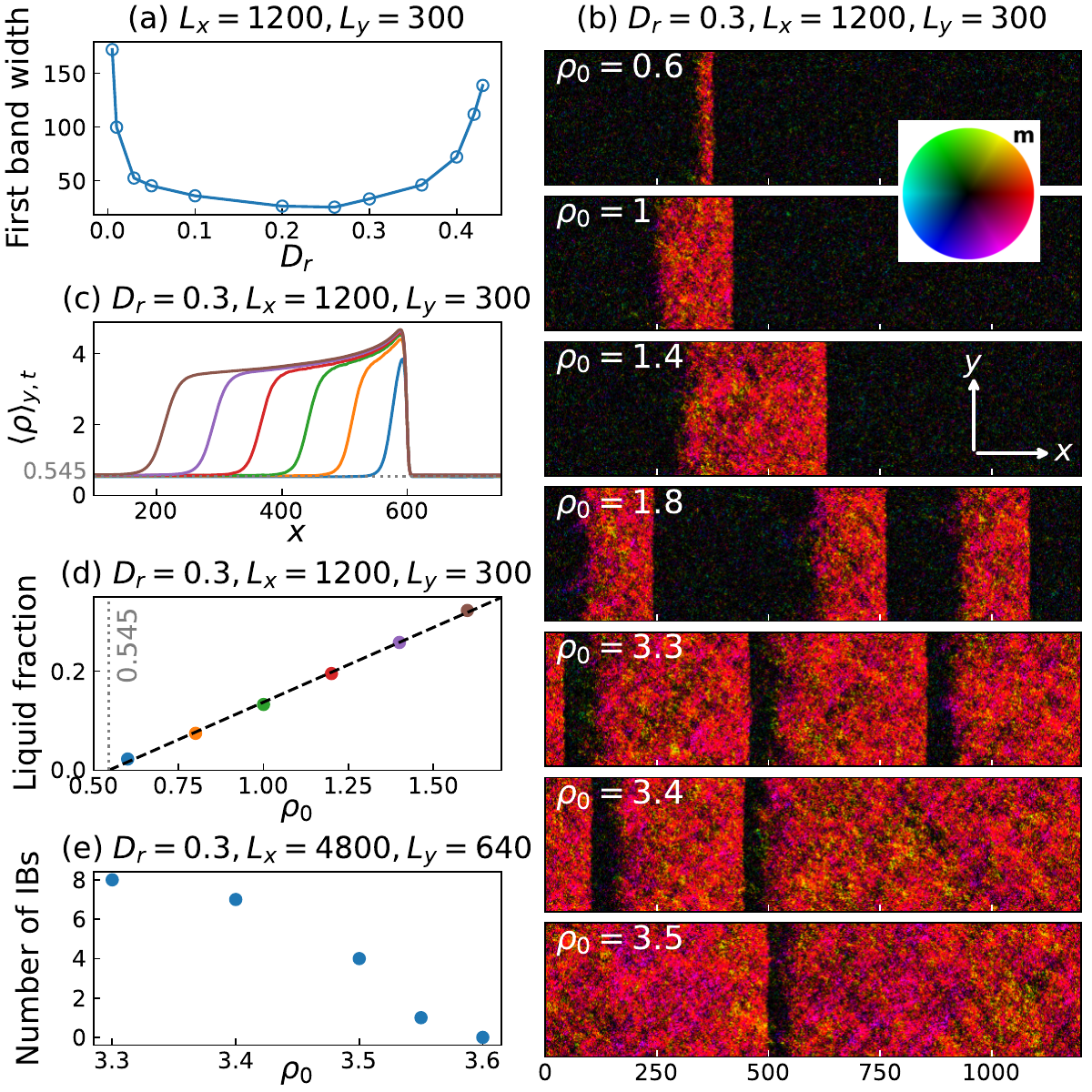}
    \caption{Regular band configurations in moderate-size systems. 
    (a) variation with $D_r$ of the width at mid height of the first band nucleating near the gas binodal (protocol
    detailed in the CP~\cite{CP}).
    (b) snapshots of the momentum field 
     in steady states obtained by increasing density slowly from $\rho_0=0.6$ to $3.5$.  
    (c) time- and $y$-averaged density profiles of the single band at different $\rho_0$ values (color coding of $\rho_0$ given in (d)).
    (d) width at mid-height of the profiles shown in (c), normalized by $L_x$: 
   the lever rule is satisfied in this apparent macrophase separation. The vertical dotted line corresponds to the gas binodal determined by the density in the gas.
    (e) decrease with increasing $\rho_0$ of the number of inverse bands.
    }
\label{fig2}
\end{figure}

Many variants of the Vicsek model have been considered over the years, with most of them equivalent 
at a qualitative level~\cite{ginelli2016physics}. 
Here we use a continuous-time version 
with standard active Brownian particles.
Position ${\bf r}_j$ and velocity orientation $\theta_j$ of particle $j$ are governed, in two space dimensions, by the following equations:
\begin{equation}
\label{VMcont}
\dot{\bf r}_j = v_0 \,{\bf e}(\theta_j) \;\;\;\; \dot{\theta_j} = \langle \sin(\theta_k - \theta_j) \rangle_k + \sqrt{2D_r} \, \xi_j \;,
\end{equation}
where $\mathbf{e}(\theta)\equiv (\cos\theta, \sin\theta)$, $\xi_j$ are white noises with zero mean and unit variance, the neighbors $k$ are particles within unit distance of $j$ and
we have, in all generality, taken a unit coupling constant in front of the interaction term.
We also set the particle speed $v_0=1$ and vary the standard main parameters, the rotational noise strength $D_r$
and the global number density of particles $\rho_0$. 
All simulations are performed on rectangular 
$L_x\times L_y$ 
domains with periodic boundary conditions, using explicit Euler-Maruyama time-stepping with $dt=0.1$.
All snapshots shown below are for the  momentum field $\mathbf{m}(\mathbf{r})=\sum_j \delta(\mathbf{r}_j-\mathbf{r})\mathbf{e}(\theta_j)$ coarse-grained over boxes of linear size 2 (see colorwheel in Fig.~\ref{fig2}(b)).

{\it Gas binodal, weak and strong noise regimes.}
The gas binodal is easily determined, following \cite{solon2015phase}. 
At fixed $D_r$, increasing $\rho_0$, a single ``first band'' is eventually nucleated. 
The binodal value is then given by reading the gas density $\rho_{gas}$.  
The blue line in Fig.~\ref{fig1}(b) was obtained this way. 
It has the expected linear behavior around the origin, and converges to the infinite-density, global coupling limit 
$D_r^\infty=\tfrac{1}{2}$
like $\rho_{gas}\sim1/(\tfrac{1}{2}-D_r)^2$.
The profile of the first band varies considerably along the binodal. 
Interestingly its width varies non-monotonously, diverging in the $D_r\to 0,\tfrac{1}{2}$ limits
and taking a minimum value near $D_r=0.23$ (Fig.~\ref{fig2}(a)). 
(See CP \cite{CP} for details and a discussion of these limits.)
The intermediate region $D_r=0.20-0.25$ roughly separates a weak noise and a strong noise regime. 

{\it Fat and inverse bands.}
Increasing $\rho_0$ at fixed $D_r$ in the weak noise regime, 
one often observes configurations
containing more and more bands parallel to the short dimension $L_y$
that are nearly identical to the first band observed near the gas binodal.
This is the microphase separation scenario described in \cite{solon2015phase}, in which the number of 
bands is proportional to $\rho_0$ and to $L_x$.

The scenario can look very different in the strong noise regime (Fig.~\ref{fig2}(b-e), details in the CP \cite{CP}): 
Increasing gradually $\rho_0$ from the gas binodal in similar rectangular geometries, 
the initial, relatively thin band gets thicker and thicker, with its width $w$ roughly proportional to $\rho_0$. 
This is typical of {\it macrophase} separation! Indeed time-averaged band profiles reveal a well-defined gas level
and the lever rule is satisfied (Fig.~\ref{fig2}(b-d)).
However, increasing $\rho_0$ further, $w$ reaches a maximal value $w_{max}$, 
and one then observes several such fat bands. 
At some point these bands cover most of the system 
and are only separated by thin holes, which we call inverse bands,
that have a well-defined structure. 
Finally, the number of inverse bands decreases to zero, leaving the ordered TT liquid. This is reminiscent
of some ``inverted'' microphase separation scenario where the elementary objects would be inverse bands, 
but no lever rule is observed (their number does not vary linearly with $\rho_0$, Fig.~\ref{fig2}(e)).
In fact, in longer domains of similar width, the number of inverse bands becomes unsteady, 
varying on long timescales, further ruining the picture of a
 microphase separation scenario (see CP \cite{CP}).

The above results are typical of systems of moderate transverse size.
Increasing it allows to consider very long fat and inverse bands by, {\it e.g.}, 
duplicating smaller systems such as those shown in Fig.~\ref{fig2}(b). 
One then observes that the trailing edge of fat bands undergoes a long-wavelength instability 
leading them to disappear by shedding away thin bands. 
Long initial inverse bands also become irregular and break, leaving spatiotemporally disordered patterns. 
(These instabilities are illustrated in the CP \cite{CP}.)

{\it Liquid binodal and super phase-separation.}
The frontier separating the coexistence region and the Toner-Tu ordered liquid 
was never defined cleanly. 
The red line in Fig.~\ref{fig1}(a) covers only rather small $\rho_0$ values and 
was determined by the location of a smooth ad-hoc crossover to the TT
liquid in systems of moderate size~\cite{chate2008collective,solon2015phase,chate2020dry}.

\begin{figure}[t!]
  \includegraphics[width=\columnwidth]{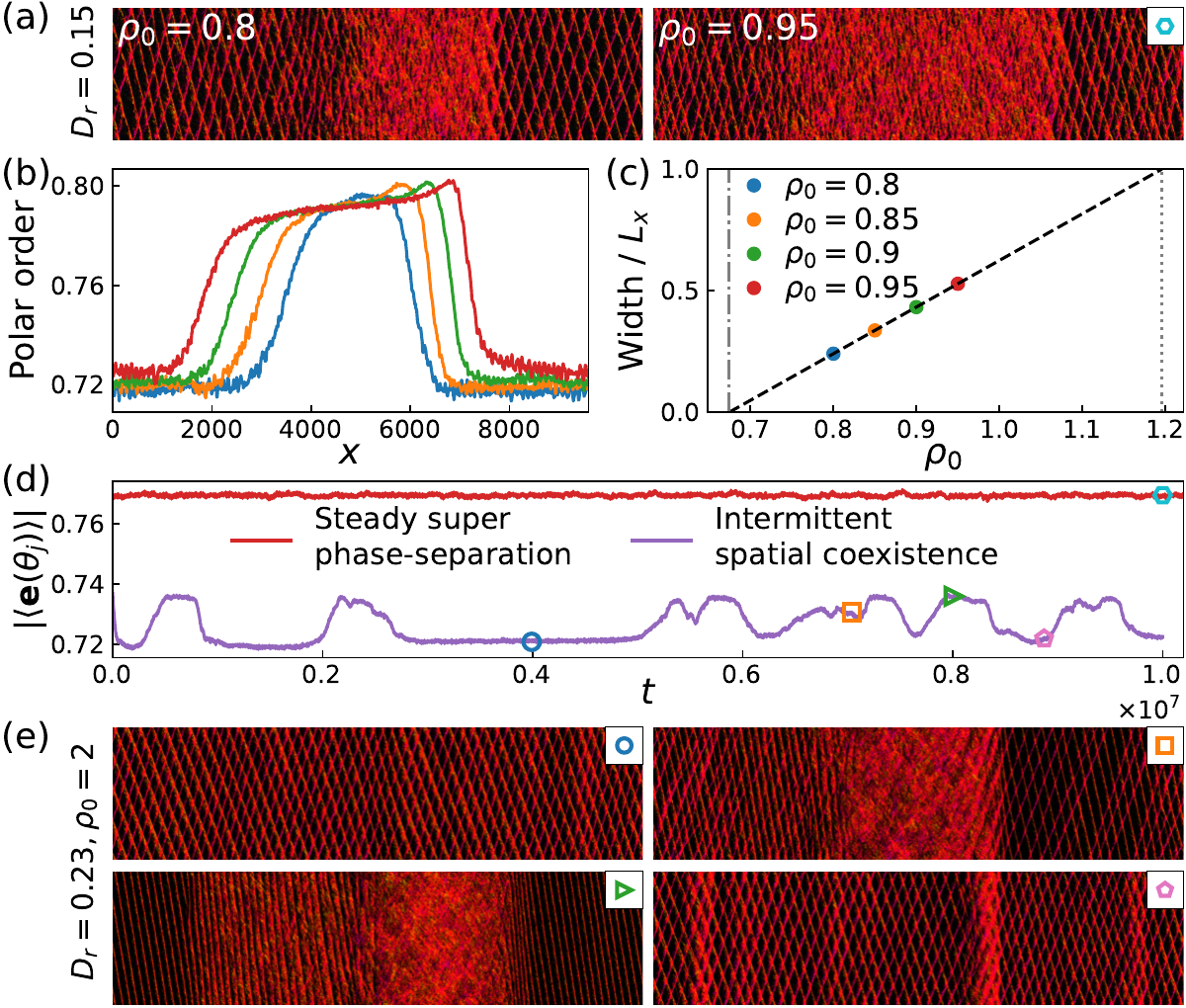}
    \caption{
    Spatial coexistence of TT liquid and bands in large systems (here $9600 \times 2400$).
    Steady coexistence at $D_r=0.15$ in (a,b,c) and red timeseries in (d), 
    intermittent coexistence at $D_r=0.23$ in (e) and purple time series in (d).
    (a) snapshots of momentum field for two different global densities. 
    (b) profiles of order averaged over the transverse direction and time $\langle \mathbf{m}(\mathbf{r},t)\cdot \mathbf{e}_x\rangle_{y, t}/\langle \rho(\mathbf{r},t)\rangle_{y,t}$ for different $\rho_0$ values indicated in (c). 
    The denser, more ordered, TT liquid domain moves faster than the cross-sea pattern of bands.
    (c) linear variation with $\rho_0$ of the longitudinal size of the TT liquid domain, measured as the width of the profiles shown in (b).
    (d) time series of global polar order parameter in steady (red line, $\rho_0=0.95$, $D_r=0.15$) 
    and intermittent (purple line, $\rho_0=2$, $D_r=0.23$) spatial coexistence of TT liquid and bands. 
    Colored symbols correspond to the snapshots shown in (a) and (e).
    (e) snapshots of momentum field taken along the timeseries shown in (d) (purple line).
      }
\label{fig3}
\end{figure}

In Figure~\ref{fig3} we show that, in very large systems,
bands and large TT liquid domains can coexist {\it in space}.
In the weak noise regime, a typically single domain of TT liquid moves steadily 
in a remaining region of crossing thin bands, which themselves move, but slower (Fig.~\ref{fig3}(a,b), Movie~1 in \cite{SUPP}). 
Increasing $\rho_0$, the fraction occupied by the liquid domain increases linearly, 
in a classic macrophase separation scenario, where the lever rule is satisfied (Fig.~\ref{fig3}(b,c)).
This leads to the definition of {\it two} binodals, the liquid binodal where the liquid fraction reaches 100\%, 
and, at lower density, the ``bands binodal'' where a large liquid domain never appears.
The red and the orange lines in Fig.~\ref{fig1}(b), which locate these two binodal lines, 
have been obtained via very large scale simulations at various values of $D_r$.

In the strong noise regime ($D_r\gtrsim0.2$), 
however, the liquid domain disappears and reappears irregularly, 
emerging from the cross-sea pattern, which it can destroy to coexist with some band smectic
(Fig.~\ref{fig3}(d,e), Movie~2 in \cite{SUPP}).
This intermittent dynamics takes place over very large timescales, 
preventing the use of the lever rule construction of Fig.~\ref{fig3}(c). 
In this intermittent phase separation regime, we located the binodals less accurately, by 
estimating when no sizeable liquid domain ever forms (bands binodal) and when the TT liquid never
leaves space to a domain of crossing bands (liquid binodal)~\footnote{
Near the liquid binodal estimated this way, some remaining inverse bands pepper the liquid.
One can track when they fully disappear, e.g. as in the ``inverse microphase scenario'' shown in Fig.~\ref{fig2}(e).
This yields a slightly different location for the liquid binodal, but we have indications that these two 
estimates converge to a single one at very large sizes.}.

Regular cross-sea patterns are observed near the bands binodal (all along the left border
of the orange line in Fig.~\ref{fig1}(b)). 
Their predominance there does not mean that they form a bona fide phase, as claimed in \cite{kursten2020dry}: 
as explained in the CP, stable configurations with crossing bands exist
throughout the region between the gas and the bands binodals. 
There is thus no cross-sea region in the new phase diagram (Fig.~\ref{fig1}(b)).

{\it Metastability of the ordered liquid.}
We finally map the region of metastability of the TT liquid, a phenomenon uncovered in 
\cite{codina2022small} and recently put on analytical ground in the high density limit~\cite{granek2026how}. 
In short: for strong-enough noise, a large-enough local perturbation triggers the growth of a
 counter-propagating, ballistically-growing droplet, 
 in striking contrast to equilibrium physics where surface tension eliminates such excitations. 
There exists, however, a noise strength below which any initial perturbation decays, regardless of its magnitude. 

\begin{figure}[t!]
  \includegraphics[width=\columnwidth]{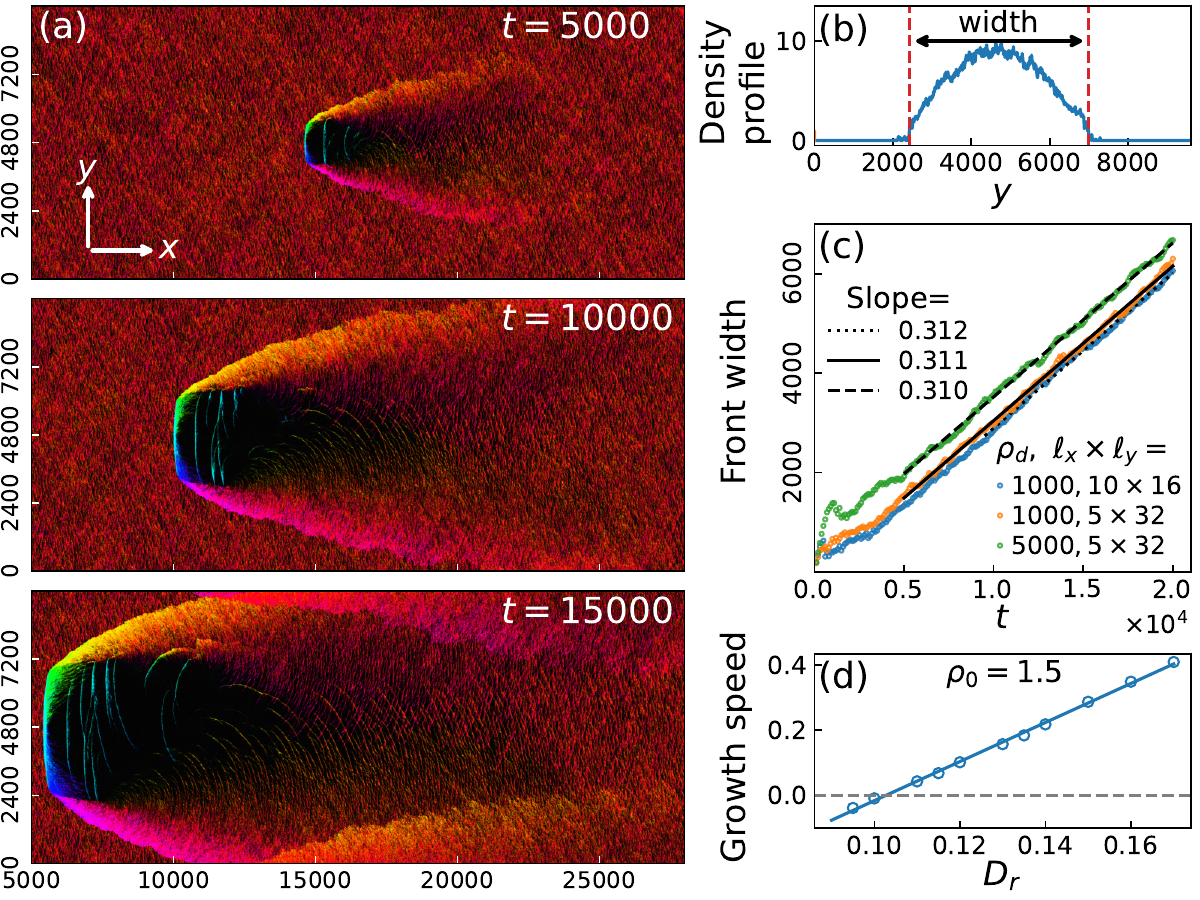}
    \caption{Metastability of the Toner-Tu liquid.
    (a) snapshots of coarse-grained momentum field during the growth of the droplet following the introduction of a group of counter-going particles ($D_r=0.15$, $\rho_0=1.2$, $L_x=38400, L_y=9600$,
    colormap as in Fig.~\ref{fig2}(b)). 
    (b) instantaneous density profile of the growing front for the $t=5000$ snapshot in (a), calculated by counting particles with $\cos\theta_i<0$ (going against the mean $+x$ flow) for each $y$ value. 
    The vertical dashed lines are
    defined by using a threshold at 10\% of the peak height. 
    (c) time series of front width (measured as in (b)) for various initial rectangular perturbations of density $\rho_d$ (colored symbols) and linear fits (black lines) over the ballistic regime revealing identical growth speed $v_g=0.311(1)$. In the ballistic regime, droplets of identical width have identical shapes (not shown).
    (d) variation of $v_g$ with $D_r$ at fixed $\rho_0=1.5$  ($L_x=38400, L_y=9600$) 
       }
\label{fig4}
\end{figure}

As in \cite{codina2022small,granek2026how}, 
we introduce a localized perturbation made of particles oriented opposite to the current global order direction  
in a steady-state ordered liquid configuration. 
The particular shape and density profile of this perturbation do not matter: 
successful, growing droplets eventually all reach the same well-defined shape that expands ballistically
at the same speed (Fig.~\ref{fig4}(a-c), Movie~3 in \cite{SUPP}). Their leading front is then essentially flat.
However, some initial shapes are better than others in the sense that 
not too many of their particles are ``lost" to the surrounding liquid, influencing the time taken to reach 
the ballistic regime.
Below, we use a rather efficient transverse rectangular domain of size $32\times 5$. 
For the cases shown in Fig.~\ref{fig4}(c), the ballistic regime is reached after approximately 5000 time units.

To estimate the asymptotic position of the metastability limit, we strived to reach the ballistic
regime so that we could estimate the front growth speed $v_g$~\footnote{The difficulty lies in ensuring that the system size is wide enough (the front width must be smaller) and long enough (the front must always visit unperturbed TT liquid).}.
As $D_r$ is decreased at a given density, $v_g$ decreases linearly,
reaching even ``negative'' values when the initial transient growth of the droplet is followed by a slow decay
 (Fig.~\ref{fig4}(d)). Leveraging this linear behavior, the metastability limit is defined as the $D_r$ value
 where $v_g=0$. 
 
 The green line of Fig.~\ref{fig1}(b) was obtained by repeating this procedure at various $\rho_0$ 
 values~\footnote{For $\rho_0>2$ this procedure becomes numerically too demanding, and we simply studied the fate of a very dense initial perturbation ($\rho_d=10^4$), locating the transition at the limit between growth and decay, without reaching the ballistic regime. For $\rho_0=1.5$, the two methods yield the same transition point.}.
It meets the liquid binodal near $\rho_0=0.4$, indicating that the TT liquid seems ``fully stable" at smaller densities.
In End Matter, we show that a phenomenological theory allows us to account accurately for this line, except at low densities.

{\it Discussion and conclusions.}
Simulations of the two-dimensional Vicsek model on length- and time-scales never explored before
have revealed that the coexistence phase separating the disordered gas
from the ordered Toner-Tu liquid is more complicated than heretofore believed, exhibiting 
new types of bands and new families of ``stable'' band configurations~\footnote{
Often, different configurations coexist at the same parameter values, and appear 
stable on the timescales probed here (from $10^6$ to $10^8$ time units).
The model being a stochastic system there cannot be coexisting attractors at finite size.
The probabilities to transit in time from one solution to the other must all be very small but non-vanishing. 
They must also be unequal so that one could in principle determine the most stable configuration from the
fraction of time spent in it in an infinitely long simulation (see further discussions in the CP~\cite{CP}).
}. 

While the concept of phase separation remains relevant, we found many declinations of it, 
different from the simple microphase separation scenario put forward in~\cite{solon2015phase}.
We showed in particular that the liquid binodal between the Toner-Tu liquid and the coexistence region
is accompanied by another binodal in an overall ``super phase separation'' scenario between the liquid and an already phase-separated phase made of crossing bands.

The  ``very large system size phase diagram'' shown in Fig.~\ref{fig1}(b) fills the gaps that motivated our study.
It is by no means definitive, but it constitutes a major update from existing knowledge.
It is in particular susceptible to evolve in the region where intermittent spatial coexistence of the Toner-Tu liquid
and crossing bands occurs ($D_r\gtrsim0.2$).
There the bands observed are often fat bands which are unstable when long enough, 
and break into multiple thin bands.

We also mapped out the limit of metastability of the Toner-Tu liquid (green line in Fig.~\ref{fig1}(b)).
In the metastability region we expect the asymptotic behavior to be some very large scale spatiotemporal band chaos,
as illustrated in \cite{codina2022small}. We nevertheless still label this region ``metastable Toner-Tu liquid"
in Fig.~\ref{fig1}(b).

As shown in the CP \cite{CP},
the band patterns emblematic of the coexistence region, smectic trains or cross-sea, can also be metastable. 
We currently do not know whether the {\it entire} coexistence region is in fact metastable. In such a scenario,
the actual asymptotic phase diagram of the model would regain a simple structure with only the blue gas binodal 
and green metastability line remaining, 
the entire region between them (shaded in grey in Fig.~\ref{fig1}(b)) then being the seat of large-scale band chaos~\footnote{ 
Note that a similar conclusion would follow if the elementary thin band solution would be proven to be linearly unstable.}. Note that this scenario is not far from what happens with the active Ising model, for which we know 
(from \cite{benvegnen2023metastability}) that
the entire liquid phase, and thus also the entire coexistence phase, are metastable.

We finally briefly discuss whether known continuous theories derived from the Vicsek model 
---most of them variants of the Toner-Tu equations---
can account for the phenomena discussed above. 
We expect many if not all of the structures uncovered here at particle level to be solutions of 
mean-field, deterministic equations~\cite{bertin2009hydrodynamic,mishra2010fluctuations,gopinath2012dynamical,ihle2013invasion,thuroff2014numerical}.
One well-established class of such equations is obtained via the Boltzmann-Ginzburg-Landau method~\cite{peshkov2014boltzmann,chate2020dry,chateDryAligningDilute2022a},
yielding basic features of the phase diagram such as 
the spinodal lines obtained from the linear stability of the homogeneous solutions.
These standard Toner-Tu equations support smectic trains of traveling bands, many of them linearly stable \cite{caussin2014emergent,solon2015pattern}, 
with a unique one selected by noise \cite{solon2015phase}.
However, to the best of our knowledge, they do not support solutions with crossing bands, 
which probably require higher-order nonlinearities. 
Similarly, as argued in the CP~\cite{CP}, solutions like the fat and inverse bands present at high density
do not fit within the dynamical system picture developed in \cite{solon2015pattern} and also require higher-order terms~\footnote{
Interesting ideas of possible relevance in this context were proposed in \cite{bertrand2022diversity}.}. 
Clearly, significant progress is needed to derive PDEs more sophisticated than the Toner-Tu equations that can
account for the large-scale phenomenology of the Vicsek model uncovered in this work.

\acknowledgments
We thank Alexandre Solon and Yongfeng Zhao for a careful reading of the manuscript.
This work is supported by the National Natural Science Foundation of China 
(Grants No.~12504239 to Y. D. and No.~12275188 to X.-q.S.); the Basic Research Program of Jiangsu (No.~BK20250788 to Y. D.); and the Natural Science Foundation of Jiangsu Province (No. BK20233001).

\bibliography{ref.bib}
\onecolumngrid

\vspace{12pt}
\noindent\hrulefill \hspace{24pt} {\bf End Matter} \hspace{24pt} \hrulefill
\vspace{12pt}

\twocolumngrid

\begin{figure*}[t]
\begin{minipage}[t]{0.48\textwidth}
    \vspace{0pt}
    \centering
    \includegraphics[width=\linewidth]{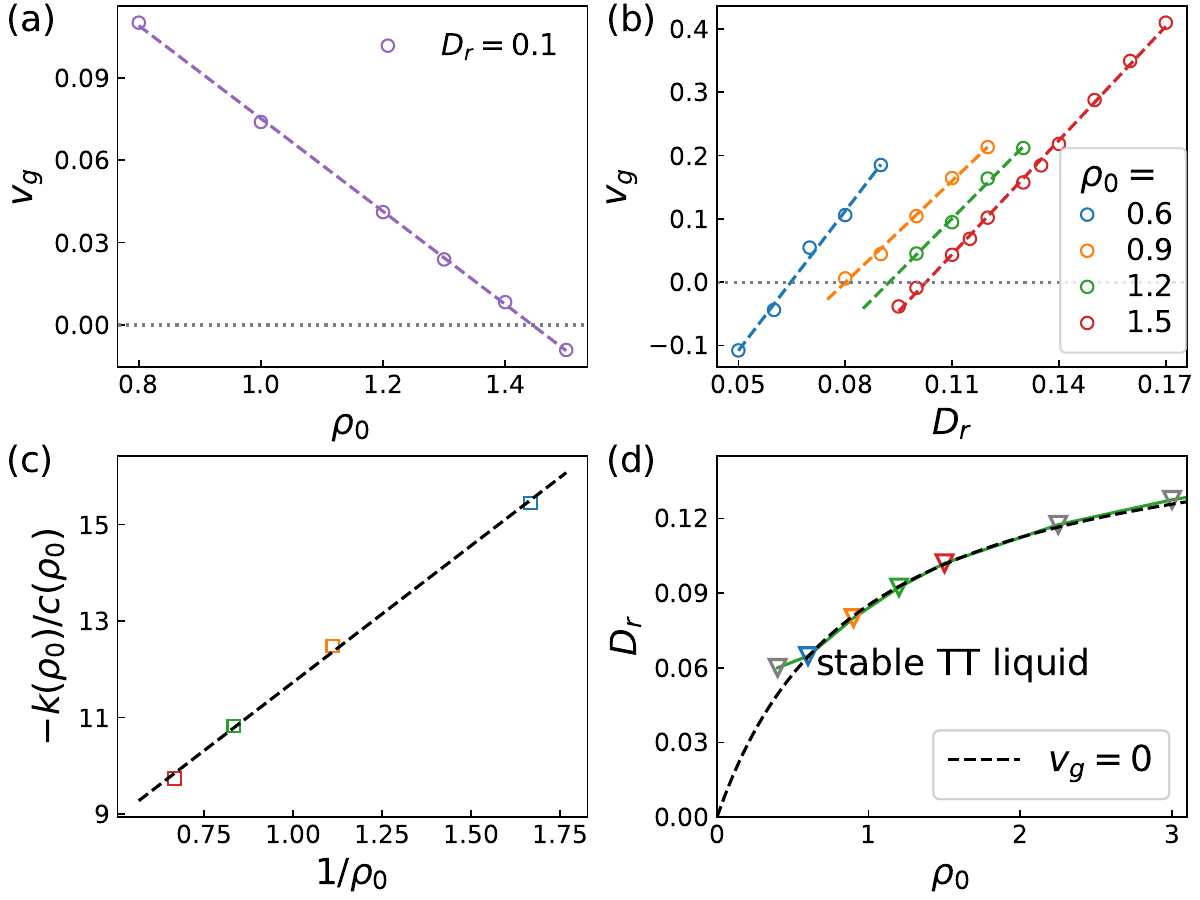}
    \caption{Metastability of the Toner-Tu liquid.
    (a) $v_g$ vs $\rho_0$ ($D_r=0.1$), measured as in Fig.~\ref{fig4}(d).
    (b) $v_g(D_r)$ for various $\rho_0$ values.
    (c) $-k(\rho_0)/c(\rho_0)$ vs $1/\rho_0$ estimated from the data shown in (b). 
    In (a-c), dashed lines are linear fits.
    (d) the $v_g=0$ dashed line given by Eq.\eqref{eqdr}, and metastability limits estimated numerically (symbols, same data as green line in Fig.~\ref{fig1}(b)).
          }
    \label{fig5}
\end{minipage}
\hfill
\begin{minipage}[t]{0.48\textwidth}
    \vspace{0pt}
    \centering
    \includegraphics[width=\linewidth]{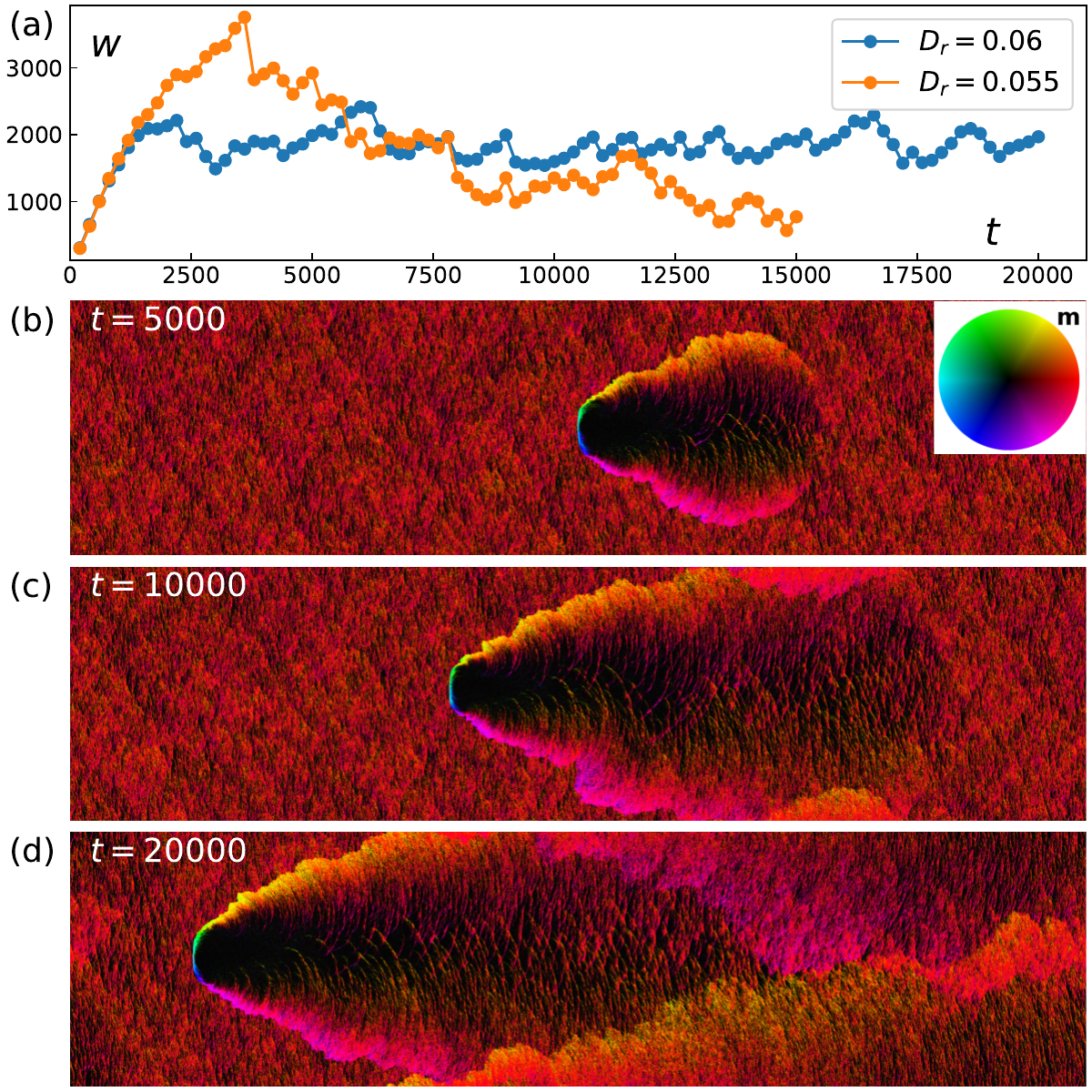}
    \caption{Metastability of the TT liquid in the low density regime 
    ($\rho_0=0.4$, $L_x=38400$, $L_y=9600$). 
    (a) time series of the droplet width for typical runs at $D_r=0.06$ and $D_r=0.055$. 
    The density of the small initial rectangular perturbation is $1.0\times10^4$ for $D_r=0.06$, and $2.5\times 10^4$ for $D_r=0.055$, which is why the orange curve overshoots the blue one.
    (b-d) snapshots of momentum field taken during the run corresponding to the blue line in (a), which
    show that the droplet does not grow nor shrink.}
    \label{fig6}
\end{minipage}
\end{figure*}

{\it Phenomenological theory for the metastability limit of the TT liquid.}
We showed in Fig.~\ref{fig4}(d) that $v_g$, the growth speed of the front width of the counter-going droplet 
expanding ballistically against the TT liquid, increases linearly with noise strength at fixed density. 
In Fig.~\ref{fig5}(a), we show that $v_g$ has also a simple dependence on density at fixed noise strength: 
$v_g$ decreases linearly with increasing $\rho_0$.

We have observed these linear variations in all of our results. This suggests the following expression for $v_g$:
\begin{equation}
\label{eqvg}
v_g(\rho_0,D_r) = \alpha D_r - \beta \rho_0 +\gamma D_r \rho_0 \;,
\end{equation}
where $\alpha$, $\beta$, and $\gamma$ are positive constants.

The term $-\beta\rho_0$ can render $v_g$  negative for large enough $\rho_0$ or small enough $D_r$, in which case any finite-size counter-propagating droplet will eventually disappear, leaving the polar liquid stable.
In this framework, the limit of metastability of the TT liquid is given by setting $v_g=0$, which, applied to  \eqref{eqvg} yields
\begin{equation}
D_r = \frac{\beta\rho_0}{\alpha+\gamma\rho_0} = \frac{\rho_0}{\alpha/\beta + (\gamma/\beta)\rho_0} \;.
\label{eqdr}
\end{equation}
The two ratios $\alpha/\beta$ and $\gamma/\beta$ can be estimated from $v_g(\rho_0,D_r)$ data such as those 
shown in Fig.~\ref{fig5}(b).
At a given $\rho_0$, we have $v_g=k(\rho_0)D_r+c(\rho_0)$, where $k(\rho_0)$ and $c(\rho_0)$ are estimated from a linear fit. From \eqref{eqvg}, we have
\begin{align}
\alpha + \gamma\rho_0 &= k(\rho_0) \label{eqk}\\
-\beta\rho_0 &= c(\rho_0) \label{eqc}
\end{align}
Dividing \eqref{eqk} by \eqref{eqc}, we obtain
\begin{equation}
\frac{\alpha}{\beta} \frac{1}{\rho_0} + \frac{\gamma}{\beta} = - \frac{k(\rho_0)}{c(\rho_0)} \;,
\end{equation}
so that $\alpha/\beta$ and $\gamma/\beta$ can be estimated from the linear dependence of $-k/c$ on $1/\rho_0$.
Using the data of Fig.~\ref{fig5}(b) for $\rho_0=0.6, 0.9, 1.2, 1.5$, we do obtain such a linear dependence (Fig.~\ref{fig5}(c)), from which we estimate
$\alpha/\beta\simeq 6.157$ and  $\gamma/\beta\simeq5.658$.

Substituting these values into \eqref{eqdr}, we obtain the dashed line in the $(\rho_0,D_r)$ plane shown in Fig.~\ref{fig5}(d), together with the data points of the numerically-estimated limit of metastability.
The agreement is very good for $\rho_0\gtrsim 0.6$, but for lower density
our phenomenological theory goes smoothly to the origin while the numerical estimates depart from it and 
meet the liquid binodal (see Fig.~\ref{fig1}(b)).

Data obtained for $\rho_0=0.4$ testify to this. 
Multiple runs performed at $D_r=0.06$, 
one of which is shown in Fig.~\ref{fig6} (blue line in (a) and snapshots below, Movie~4 in \cite{SUPP}), 
all produced a droplet whose width,
after some initial growth, fluctuates around a constant value. 
In contrast, the same experiments performed at $D_r=0.055$ all lead to a slowly disappearing droplet 
(orange line in Fig.~\ref{fig6}(a)). This implies that, at this density, the $v_g=0$ point is very near $D_r=0.06$
and definitely above $D_r=0.05$, the value given by the phenomenological theory. 

We trace the origin of this to the stronger fluctuations present in the TT liquid at low densities 
(compare the TT liquid in Fig.~\ref{fig6}(b-d) and in Fig.~\ref{fig4}(a)).  
The growth speed $v_g$ is ultimately set by the competition 
of the particle fluxes flowing in and out of the front of the counter-propagating droplet at its two edges. 
We believe these fluxes are influenced by the stronger polar clusters forming fluctuations in the liquid,
something not well taken into account by our simple Eq.~\eqref{eqvg}. 

\end{document}


\begin{CJK*}{UTF8}{gbsn}

\title{Supplemental Material to ``Phase Diagram of the Two-Dimensional Vicsek Model"}

\author{Yu Duan ({\CJKfamily{gbsn}段煜})}
\affiliation{Center for Soft Condensed Matter Physics and Interdisciplinary Research \& School of Physical Science and Technology, Soochow University, Suzhou 215006, China}
\affiliation{Max Planck Institute for Dynamics and Self-Organization (MPI-DS), 37077 G\"ottingen, Germany}
\affiliation{Jiangsu Physical Science Research Center, Nanjing 210093, China}

\author{Hugues Chat\'{e}}
\affiliation{Service de Physique de l'Etat Condens\'e, CEA, CNRS Universit\'e Paris-Saclay, CEA-Saclay, 91191 Gif-sur-Yvette, France}
\affiliation{Sorbonne Universit\'e, CNRS, Laboratoire de Physique Th\'eorique de la Mati\`ere Condens\'ee, 75005 Paris, France}

\author{Beno\^it Mahault}
\affiliation{Laboratoire Charles Coulomb, Universit\'e de Montpellier, CNRS, Montpellier, France}
\affiliation{Max Planck Institute for Dynamics and Self-Organization (MPI-DS), 37077 G\"ottingen, Germany}

\author{Xia-qing Shi ({\CJKfamily{gbsn}施夏清})}
\affiliation{Center for Soft Condensed Matter Physics and Interdisciplinary Research \& School of Physical Science and Technology, Soochow University, Suzhou 215006, China}

\maketitle
\end{CJK*}

\section{DESCRIPTION OF THE SUPPLEMENTAL MOVIES}
\begin{itemize}
    \item \texttt{Movie\_1.mp4}. Steady super coexistence between bands and TT liquid, corresponding to the right panel of Fig.~3(a) of the main text. Parameters: $L_x=4 L_y=9600$, $D_r=0.15$, $\rho_0=0.95$.
    \item \texttt{Movie\_2.mp4}. Intermittent spatial coexistence between bands and TT liquid, corresponding to Fig.~3(e) of the main text. Parameters: $L_x=4 L_y=9600$, $D_r=0.23$, $\rho_0=2.0$.
    \item \texttt{Movie\_3.mp4}. Growing counter-propagating front, corresponding Fig.~4(a) of the main text. The initial perturbation is introduced using a rectangular domain of size $(\ell_x=5, \ell_y=32)$, in which particles of density $1.0\times 10^3$ oriented along the $-x$ axis, opposite to the initial global order direction. Parameters: $L_x=4L_y=38400$, $D_r=0.15$, $\rho_0=1.2$
    \item \texttt{Movie\_4.mp4}. Counter-propagating front with growth speed $\approx 0$, corresponding to the blue curve in Fig.~6(a) and snapshots in Fig.~6(b-d) of the main text. The initial rectangular perturbation has size $(\ell_x=5, \ell_y=32)$ and density $1.0\times 10^4$. Parameters: $L_x=4L_y=38400$, $D_r=0.06$, $\rho_0=0.4$
\end{itemize}